\documentclass[12pt,showpacs,showkeys]{revtex4}
\usepackage{amssymb,amsmath}

\begin{document}

\title{Electric/magnetic reciprocity in premetric electrodynamics with
  and without magnetic charge, and the complex electromagnetic field}

\author{Friedrich W. Hehl}
\email{hehl@thp.uni-koeln.de}

\affiliation{Inst.\ Theor.\ Physics, University of Cologne, 50923
  K\"oln, Germany \& Dept.\ Physics and Astronomy, University of
  Missouri-Columbia, Columbia, MO 65211, USA}

\author{Yuri N. Obukhov}
\email{yo@thp.uni-koeln.de}

\affiliation{Inst.\ Theor.\ Physics, University of Cologne, 50923
  K\"oln, Germany \& Dept.\ Theor.\ Physics, Moscow State University,
  117234 Moscow, Russia}

\date{09 Feb 2004, {\it file recip9a.tex}}

\begin{abstract}
  We extend an axiomatic approach to classical electrodynamics, which
  we developed recently, to the case of non-vanishing magnetic charge.
  Then two axioms, namely those of the existence of the Lorentz force
  (Axiom 2) and of magnetic flux conservation (Axiom 3) have to be
  generalized.  Electric/magnetic reciprocity constitutes a guiding
  principle for this undertaking. The extension of the axioms can be
  implemented at a premetric stage, i.e., when metric and connection
  of spacetime don't play a role. Complex Riemann-Silberstein fields
  of the form $(E\pm i\,{\mathcal H},{\mathcal D}\pm i\,B)$ have a
  natural place in the theory, {\it independent} of the Hodge duality
  mapping defined by any particular metric.
\end{abstract}

\maketitle

\newpage

\section{Magnetic charge and electric/magnetic reciprocity}
\label{12axioms}

An axiomatic approach to classical electrodynamics without magnetic
charges (mono\-poles) has been developed in \cite{HO3} (see also \cite{Post}
and \cite{TT60}). In this approach, the initial structure of spacetime is 
that of a 4-dimensional differentiable manifold foliated by means of a
monotonically increasing `topological time' parameter $\sigma$ (see
also \cite{Oz}). Maxwell's equations are expressed in terms of the
twisted \it excitation \rm 2-form $H$, the twisted electric current
3-form $J$, and the untwisted \it field strength \rm 2-form $F$:
\begin{equation}\label{Maxwell}
dH=J\,,\qquad dF = 0\,.
\end{equation}
The different fields decompose into time and space pieces according to 
\begin{eqnarray}\label{pieces}
  H &=& -\,{\mathcal H}\wedge d\sigma+ {\mathcal D}\,, \quad F=E\wedge
  d\sigma+ B\,,\\ J &=& -\,j\wedge d\sigma+\rho\,,\label{pieces'}
\end{eqnarray}
where, in 3 dimensions, ${\mathcal H}$ and $E$ are 1-forms, ${\mathcal
  D}$, $B$, and $j$ are 2-forms, and $\rho$ is a 3-form.
Eqs.(\ref{pieces}) and (\ref{pieces'}) provide the physical
interpretation of the 4-dimensional diffeomorphism and frame invariant
Maxwell equations (\ref{Maxwell}) in terms of measurable quantities,
namely the electric and the magnetic excitations ${\mathcal
  D},{\mathcal H}$, the electric and magnetic field strengths $E,B$,
and the electric charge density $\rho$ and its current density $j$,
respectively. Up to here, we are still on a {\it pre\/}metric level:
Neither a metric of spacetime nor a connection are assumed to exist.
Additional spacetime structures, including the light cone and other
elements of a Lorentzian metric, emerge only after a further condition
is imposed via a spacetime relation defining the electromagnetic
properties of spacetime (of the `vacuum') \cite{Eldyn1} (see also
\cite{IH3}).

In Maxwell's theory as well as in our axiomatics a clear asymmetry is
built in between electric and magnetic charges. Magnetic charges have
been a matter of numerous theoretical discussions, see, e.g.,
Schwinger \cite{S75,S98}, and of experimental investigations. No
experimental evidence exists for magnetic charges or monopoles, see He
\cite{He}, Abbott et al.\ \cite{Abbott}, and Kalbfleisch et
al.\ \cite{Kalbfleisch}. Nevertheless, in this note, following a
discussion with Kaiser, see \cite{K04}, we want to address the
question how our axiomatics can be adapted to the introduction of
magnetic charges.

Let us remind ourselves that for the derivation of Maxwell's equations
(\ref{Maxwell}) we need Axiom 1, namely charge conservation $dJ=0$,
Axiom 2, the existence of the Lorentz force density
\begin{equation}\label{lorentz}
f_\alpha=(e_\alpha\rfloor F)\wedge J\,
\end{equation}
(here $e_\alpha$ is the frame or vierbein field), and Axiom 3, magnetic
flux conservation $dF=0$. Furthermore, for our theoretical analysis
Axiom 4 is required, the (still premetric) localization of
electromagnetic energy and momentum by means of the twisted kinematic
energy-momentum current 3-form
\begin{equation}\label{simax}
  {}^{\rm k} \Sigma_\alpha :={\frac 1 2}\left[F\wedge(e_\alpha\rfloor
    H) - H\wedge (e_\alpha\rfloor F)\right]\,.
\end{equation}
As shown in \cite{HO3}, this current, in contrast to Maxwell's
equations themselves, is {\it electric/mag\-netic reciprocal}, i.e., it
is invariant under the substitutions
\begin{equation}\label{duality1a} 
 \; H\rightarrow \!\zeta F \qquad\hspace{10pt}
\begin{cases}{\mathcal H}\rightarrow
    -\zeta E\,,\\ {\mathcal D}\rightarrow \quad\zeta B\,,
\end{cases}
\end{equation}
\begin{equation}\label{duality1b} 
  F\rightarrow -\frac{1}{\zeta}\,H
  \qquad\begin{cases}E\rightarrow \quad\!\!\frac{1}{\zeta}\,{\mathcal
      H}\,,\\ B\rightarrow -\frac{1}{\zeta}\,{\mathcal
      D}\,,\end{cases}
\end{equation} 
with a constant twisted 0-form $\zeta$ of the dimension of an
admittance. In (\ref{duality1a}), (\ref{duality1b}), a magnetic
quantity is replaced by an electric one {\em and} an electric quantity
by a magnetic one.  Accordingly, we can speak of an electric/magnetic
reciprocity of the energy-mo\-men\-tum current $^{\rm
  k}\Sigma_\alpha$. Observe the non-trivial minus sign on the
left-hand-side of (\ref{duality1b}) which is necessary because of the
structure of $^{\rm k}\Sigma_\alpha$.

In a straightforward way, we can try to introduce an untwisted
magnetic current 3-form $K$ that is conserved according to $dK=0$, in
analogy to Axiom 1, and that again could be verified just by counting
the monopoles. If this new Axiom $3'$ is suitably formulated, we would
find, by de Rham's theorem, the magnetic current $K$, analogously to
the electric current $J$, as exact 3-form: $dF=K$.  If we uphold
Axioms 1, 2, and 4, then we expect an {\it inconsistency} since,
according to Axiom 2, the electromagnetic field would only act on {\it
  electric} not, however, on magnetic charges.

Before we continue this line of arguments, let us take a look at the
`new' Maxwell equations:
\begin{equation}\label{newmaxwell}
dH=J\,,\qquad dF=K\,.
\end{equation}
If we now introduce, in addition to (\ref{duality1a}), (\ref{duality1b}),
a new substitution for the electric and magnetic currents
\begin{equation}\label{newsubs}
  J\rightarrow \zeta\,K\,,\qquad K\rightarrow -\frac{1}{\zeta}\,J\,,
\end{equation}
then not only the energy-momentum current (\ref{simax}) is
electric/magnetic reciprocal but also the Maxwell equations
(\ref{newmaxwell}) themselves.

Let us come back to the last but one paragraph. In order to remove the
inconsistency mentioned, we have, in addition to Axiom 3, also to
change at least one other axiom. As argued, Axiom 2 would now appear
to be shaky.  This axiom of standard (monopole-free) electrodynamics
defines the electromagnetic field strength $F$ by means of the Lorentz
force density acting on a moving electric charge density. The
assumption of the existence of magnetic charge requires to modify the
Lorentz force axiom correspondingly by also including a piece for the
force acting on the magnetic charge.  However, magnetic monopoles were
never observed, and thus we cannot appeal to experiment in formulating
the modification of Axiom 2. As a matter of fact, there exists quite
an ambiguity in choosing the form of the new ``Lorentz" force density,
and we can only rely on theoretical arguments.  Among other
possibilities, the use of electric/magnetic reciprocity as a guiding
principle provides a self-consistent and mathematically nice
framework.

If we supplement, in accordance with (\ref{duality1a}),
(\ref{duality1b}), and (\ref{newsubs}), the right-hand-side of
(\ref{lorentz}) such that it becomes electric/magnetic reciprocal, we
find
\begin{equation}\label{inter1}
  f_\alpha = (e_\alpha\rfloor F)\wedge J -  (e_\alpha\rfloor
  H)\wedge K\,.
\end{equation}
The second term describes the force acting on a magnetic charge. Then,
following along the same line of thought as in our standard premetric
axiomatics \cite{HO3}, we naturally arrive at the {\it unchanged}
Axiom 4 for the electromagnetic energy-momentum current, as we
specified it in (\ref{simax}).  Accordingly, by assuming the existence
of hypothetical magnetic charges, one can develop a straightforward
generalization of the axiomatic premetric approach by stressing
specifically the notion of {\it electric/magnetic reciprocity}.  The
latter can be considered as a premetric counterpart of
electric-magnetic duality.

Incidentally, the density of the magnetic mono\-pole force
$-(e_\alpha\rfloor H) \wedge K$ is not unprecedented: For magnetic (north
and/or south) poles, a Coulomb like law can be formulated and
experimentally verified to some approximation, as was done first by
Coulomb himself, see \cite{BSch66}.  Then the force acting on the unit
pole defines the magnetic excitation $\mathcal H$ in much the same way as
the electric (monopole) force defines the electric field strength $E$.
Accordingly, we assume that the hypothetical magnetic monopoles behave
in an analogous way as the effective magnetic poles of conventional
electrodynamics. Needless to say that a magnetic north pole cannot be
separated from its south pole. Therefore the magnetic pole concept
doesn't survive on a fundamental level.

In order to verify (\ref{inter1}), we substitute (\ref{newmaxwell}) in
(\ref{inter1}):
\begin{equation}\label{inter1a}
  f_\alpha = (e_\alpha\rfloor F)\wedge dH -  (e_\alpha\rfloor
  H)\wedge dF\,.
\end{equation}
This is eq.(B.5.1) of ref.\cite{HO3} on page 163 in the case of
$a=1$.  After some simple algebra, which is spelled out in \cite{HO3},
we arrive at eq.(B.5.4), loc.cit.\ (with $a=1$):
\begin{eqnarray}\label{inter3a}
  f_\alpha= d\bigl[F\wedge(e_\alpha\rfloor H)&-&H\wedge(e_\alpha\rfloor
  F)\bigr]\nonumber\\ -F\wedge ({\hbox{\it \char'44}}_{e_\alpha} H)&+&
H\wedge ({\hbox{\it \char'44}}_{e_\alpha} F)\nonumber\\ 
-(e_\alpha\rfloor F)\wedge
    dH &+& (e_\alpha\rfloor H)\wedge dF \,.
\end{eqnarray}
We denote by ${\hbox{\it \char'44}}_{e_\alpha}$ the Lie derivative with
respect to the frame $e_\alpha$.  We can substitute (\ref{simax}) in
(\ref{inter3a}). Then, with the definition of the force
density that is left over,
\begin{equation}  X_\alpha := -{\frac 1 2}\left(F\wedge
{\hbox{\it \char'44}}_{e_\alpha}H-H\wedge{\hbox{\it \char'44}}_{e_\alpha}
F \right)\,,\label{Xal}
\end{equation}
we find
\begin{eqnarray}\label{qwe}
  f_\alpha&=&2\,d\,\Sigma_\alpha+2\,X_\alpha\nonumber\\&&-
  (e_\alpha\rfloor F)\wedge dH + (e_\alpha\rfloor H)\wedge dF\,.
\end{eqnarray}
The last two terms in (\ref{qwe}) add up to $-f_\alpha$, see
(\ref{inter1}). Thus,
\begin{equation}\label{qwe'}
  f_\alpha=(e_\alpha\rfloor F)\wedge J - (e_\alpha\rfloor H)\wedge K
  =d\,\Sigma_\alpha+X_\alpha\,.
\end{equation}
This conclusion coincides with an earlier result of Kaiser \cite{K04}.

We are now coming to the central point: Note that all the expressions
in Axiom $2'$, see (\ref{inter1}), and in Axiom 4, see (\ref{simax}),
are invariant under the {\it electric/magnetic reciprocity mapping}
\rm $\circledast$ defined by
\begin{eqnarray}\label{recipr}
  \circledast H&=&\zeta F\,,\quad \circledast F=-\frac{1}{\zeta}\,H
  \,,\\ \circledast J&=&\zeta K\,,\quad \circledast
  K=-\frac{1}{\zeta}\,J\,.
\end{eqnarray}
Since $F$ has the dimension of an \it action \rm and $H$ that of a
\it charge, \rm $\zeta$ has the dimension of an \it admittance \rm or
reciprocal impedance.  Furthermore, to keep $\circledast H$ twisted
and $\circledast F$ twist-free, $\zeta$ must transform as a \it
pseudoscalar, \rm changing sign under orientation-reversing
diffeomorphisms.  Of course, by reciprocity, Axiom 1 is mapped to
Axiom 3, and vice versa. Accordingly, as we saw already, the `new'
Maxwell equations are also electric/magnetic reciprocal. In the
absence of a metric, reciprocity precedes \it electric-magnetic
duality\rm, as we will see below.  In the generalized Lorentz force
density (\ref{inter1}), the excitation $H$ (generated by the electric
current $J$ via $dH=J$) yields a force on the magnetic current $K$ via
$-(e_\alpha\rfloor H)\wedge K$. The magnetic current $K$ generates the
field strength $F$ (via $dF=K$) and $F$, in turn, yields the force
$(e_\alpha\rfloor F)\wedge J$ on $J$.  This closes the circle.

It is interesting to look into the physical meaning of the modified
Axiom 2 and Axiom 3, namely (\ref{inter1}) and (\ref{newmaxwell})$_2$.
In simple terms, they tell us that both electromagnetic fields, $H$ as
well as $F$, are simultaneously of the type of excitations (i.e.,
quantities) {\it and} of field strengths (i.e., intensities).  Thus,
if magnetic charges exist, the electric/magnetic reciprocity washes
out the difference between these notions and we naturally come to the
conclusion that these fields might be combined into a single
electric/magnetically symmetric object.

\section{Complex Structures}

We know that the expressions in (\ref{newmaxwell}), 
(\ref{inter1}), and (\ref{simax}) are invariant under the electric/magnetic 
reciprocity mapping (\ref{recipr}). This suggests that we introduce the 
complex electromagnetic field and its conjugate complex \cite{HO3}:
\begin{equation}\label{fieldU}
  U := H + i\zeta F\,,\qquad \overline{U} = H - i\zeta F\,.
\end{equation}
Then the reciprocity mapping yields
\begin{equation}\label{reciU}
\circledast  U = -iU\,,\qquad \circledast\overline{U}=+i\overline{U}\,,
\end{equation}
that is, $U$ and $\overline{U}$ are {\it eigenvectors\/} of the
reciprocity mapping and acquire a nontrivial meaning thereby. In
physical terms this means that $U$ is, up to a factor,
electric/magnetic reciprocal.  Furthermore, by implication,
$\circledast \circledast U=-U$ and $\circledast \circledast
\overline{U}=- \overline{U}$. Therefore the reciprocity mapping
induces locally an \it almost complex structure \rm on the underlying
spacetime manifold.

If we introduce the complex electric/magnetic current 
\begin{equation}\label{current}
{\mathfrak J}:= J+i\zeta K\,,
\end{equation}
then we can display the complex Maxwell equation as
\begin{equation}\label{MaxU}
  \boxed{dU=\mathfrak{J}.}
\end{equation}
If and only if we have a nontrivial magnetic current $K\ne 0$,
electric/magnetic reciprocity extends to the Maxwell equation, since
\begin{equation}\label{reciJ}
  \circledast \mathfrak{J}=-i\mathfrak{J}\,.
\end{equation}
However, conventionally, with vanishing magnetic current $K=0$, reciprocity 
is confined to the energy-momentum 3-form of the electromagnetic field (see 
(\ref{simax})), where we observed it in the first place:
\begin{eqnarray}
\Sigma_\alpha(U) &=& \frac{i}{4\zeta}\left(\overline{U}\wedge e_\alpha
\rfloor U -U\wedge e_\alpha\rfloor\overline{U} \right)\,,\nonumber\\ 
\qquad\circledast\Sigma_\alpha(U) &=& \Sigma_\alpha(U)\,.\label{emU}
\end{eqnarray}

Our earlier results are already incorporated in this generally
covariant and premetric framework developed so far in this section. To
make them explicitly visible, we can use a foliation and the
``topological'' time $\sigma$. Quite generally, a 2-form decomposes
according to
\begin{equation}\label{decompose}
  U =-U_{\bot}\wedge d\sigma +\underline{U}\,,
\end{equation}
see \cite{HO3}. Here $U_\bot$, a 1-form in 3 dimensions, is the
timelike (or longitudinal) piece of $U$ and $\underline{U}$, a 2-form,
the spacelike (or transversal) piece of $U$. By means of
(\ref{pieces}) and (\ref{decompose}), we find
\begin{equation}\label{underbot}
  U_\bot={\mathcal H}-i\zeta E\,, \qquad 
  \underline{U}= {\mathcal D} +i\zeta B\,.
\end{equation}
Thus, the {\em time\/}like component of $U$, that is, $U_\bot$, is,
apart from the factor $-i$, the Riemann-Silberstein field
$\zeta{E}+i{\mathcal H}$ and the {\em space\/}like component
$\underline{U}$ is ${\mathcal D}+i\zeta{B}$.  Accordingly, we can
decompose the Maxwell equation (\ref{MaxU}) by the method quoted. We
find,
\begin{equation}\label{MaxDec}
\underline{\dot{U}}-\underline{d}U_\bot=\mathfrak{J}_\bot\,,\qquad
\underline{d}\,\underline{U}=\underline{\mathfrak{J}}\,.
\end{equation}
Here the dot is defined by $\dot{()}:=\pounds_{n}$, where $n$ denotes
the normal vector of the spacetime folio with respect to which the Lie
derivative is taken.  In this way, we have separated the evolution
equation (\ref{MaxDec})$_1$ from the constraint equation
(\ref{MaxDec})$_2$; and the extensive quantity the time evolution of
which we are studying is the Riemann-Silberstein field ${\mathcal
  D}+i\zeta{B}$.

Complex combination of the type ${E}\pm i\,{\mathcal H}$ and 
${\mathcal D}\pm i\,{B}$ have appeared in the literature for many years. 
Bialynicki-Birula \cite{B96} has traced the occurrence of such
combinations back to Riemann and Silberstein, thus he calls them \it
Riemann-Silberstein fields. \rm They have been rediscovered many times
and applied to a wide variety of electromagnetic phenomena. Some
examples which we are familiar with include the work of Robinson
on null fields \cite{R61}, Trautman on analytic solutions of
Lorentz-invariant equations \cite{T62,Trautman}, Newman's
interpretation of the analytically continued Coulomb field as the
electromagnetic part of the Kerr-Newman solution \cite{N73,K1,N2},
Mashhoon's treatment of wave propagation in a gravitational background
\cite{M73,M74,M75}, Kaiser's construction of electromagnetic
wavelets \cite{K3}, and Bialynicki-Birula's work on the photon wave
function \cite{B96} and electromagnetic vortices \cite{BB3,B4,K4}.

The ``exterior square'' of the complex electromagnetic field
\begin{equation}\label{zymotisch}
  U\wedge U= H\wedge H
  -\zeta^2 F\wedge F +2i\zeta H\wedge F\,,
\end{equation}
if decomposed according to (\ref{pieces}),
\begin{equation}\label{UU1}
  U\wedge U= 2 d\sigma\wedge ({\mathcal D}+i\zeta B)\wedge({\mathcal H}
  -i\zeta E)\,,
\end{equation}
contains both Riemann-Silberstein fields in a symmetric form.
The premetric modulus of $U$ turns out to be
\begin{eqnarray}\label{UU*}
  \overline{U} \wedge U& =& H\wedge H +\zeta^2 F\wedge F\nonumber\\&=&
  2 d\sigma\wedge ({\mathcal H}\wedge {\mathcal D} - \zeta^2 B\wedge
  E)\,.
\end{eqnarray}

Let us stress that all the equations (\ref{Maxwell}) to (\ref{UU*}) are
generally covariant under coordinate and frame transformations. There
is no metric involved and no Poincar\'e group, but rather the
diffeomorphism group and the local linear group alone. In other words,
this is really a {\em premetric\/} approach to Maxwell's theory.

Let us now turn to {\it Maxwell-Lorentz electrodynamics} (``linear
vacuum electrodynamics'') by postulating Axiom 5, that is, the spacetime
relation
\begin{equation}\label{ML}
 H=\zeta\, ^\star F\,.
\end{equation}
The spacetime metric comes in with the Hodge star operator $\star$.
In SI (the International System of Units), the vacuum admittance is
$\zeta=\sqrt{\varepsilon_0/\mu_0}$ and the speed of light
$c=1/\sqrt{\varepsilon_0\mu_0}$. Furthermore $\sigma=t$, with the
metric time $t$.  Then we find
\begin{equation}
  U=\zeta\,(^\star F+iF)\,,\qquad \overline{U}=\zeta\,(^\star F-iF)\,,
\end{equation}
which reduce to the Hodge eigenvectors
\begin{align}
  ^\star U = iU\,,\qquad ^\star\overline{U}=-i\overline{U}\,,
\end{align}
compare (\ref{reciU}). Clearly, $^{\star\star}U=-U$ and $^{\star\star}
\overline{U}=-\overline{U}$. In this way, the Hodge star operator
inherits the almost complex structure from our electric/magnetic
reciprocity mapping. Reciprocity is primary, Hodge duality secondary.
If $^{\underline{\star}}$ is the 3-dimensional Hodge star, then we can
decompose (\ref{ML}) as
\begin{equation}\label{ML1+3}
  {\mathcal D}=\frac{\zeta}{c}\,^{\underline{\star}}E=\varepsilon_0 
\,^{\underline{\star}}E\,,\;\;
  B=\frac{1}{c\,\zeta}\,^{\underline{\star}}{\mathcal H}=
 \mu_0\,^{\underline{\star}}{\mathcal H}\,.
\end{equation}

The exterior square $U\wedge U$ of (\ref{zymotisch}), in the {\bf
  M}axwell-{\bf L}orentz case, reduces to
\begin{equation}\label{UUvac}
U\wedge U\stackrel{\rm ML}{=}-2\zeta^2\left(F\wedge
  F-i\,^\star F\wedge F \right)\,.
\end{equation}
In the Lagrangian, $F\wedge F$ would be a surface term and $^\star
F\wedge F$ be proportional to the Maxwell Lagrangian. Thus $U\wedge U$
qualifies as a Lagrangian. It can also be expressed
as $U\wedge U\stackrel{\rm ML}{=}-i\,^\star U\wedge U$. For better
insight, we can still decompose (\ref{UUvac}) into 1+3. Then
(\ref{UU1}) translates into
\begin{eqnarray}\label{sonstwas}
  U\wedge U\stackrel{\rm ML}{=} 2
  i\,dt\wedge\,^{\underline{\star}}({\mathcal H}-i\zeta E)\wedge 
 ({\mathcal H}-i\zeta E) \,.
\end{eqnarray}
Hence the square of $U$ is closely linked to the modulus of the
Riemann-Silberstein covector ${\mathcal H}-i\zeta E$.  The Minkowski space
analog of (\ref{sonstwas}) is the well-known quantity $\psi :=
(\zeta\mathbf{E}+i\mathbf{H}) \cdot(\zeta\mathbf{E} +i\mathbf{H})$. It
is, of course, also Poincar\'e-invariant. Among other things, it plays the
key role of polarization scalar in the definition of electromagnetic
vortices \cite{BB3,B4,K4}.

We substitute (\ref{ML}) into (\ref{UU*}) and find for the modulus of
$U$:
\begin{eqnarray}
  \overline{U}\wedge U&\stackrel{\rm ML}{=}&\zeta^2\left(^\star
    F\wedge{}^\star F+F\wedge F \right)\nonumber\\ 
  &=& \zeta^2\left(^{\star\star}F\wedge F+F\wedge F\right)=0\,.
\end{eqnarray}
Thus, in Minkowski space --- and, more generally, in all
pseudo-Riemannian spacetimes --- the 4-form $\overline{U}\wedge U$
vanishes.

\section{Discussion and conclusion}

In this paper, we formulated an extension of the axiomatic premetric
approach to the case of magnetic charges. Electric/magnetic
reciprocity plays a central role in such an extension. It is
instructive to check the absolute dimensions of the physical
quantities involved. Let us recall \cite{HO3} that the absolute
dimension of the electric current is that of electric charge, $[J] =
q$. As a result of Axiom 1 and Axiom 2, the absolute dimensions of the
electromagnetic field 2-forms are $[H] = q$ and $[F] = h/q$, i.e.,
electric charge and action per charge, respectively.  Now, looking at
the modified Lorentz force density (\ref{inter1}), we conclude that
the absolute dimension of the magnetic current 3-form is that of
action per charge: $[K] = h/q$. Thus the dimension of magnetic charge
turns out to be fixed. As a result classical electrodynamics may only
have 4 or less independent dimensional units (contrary to Sommerfeld
\cite{Sommerfeld}, e.g., who operated with 5 units). From a
dimensional point of view, the reciprocity operator maps electric
charge into action per charge, and vice versa. This fact obviously
underlies the well-known charge quantization which arises naturally in
the framework of an electrodynamics that includes magnetic
monopoles.

Denoting the elementary electric and magnetic charges by $e$ and $g$,
respectively, we conclude that the 2-forms $eF$ and $gH$, as well as
the 3-forms $eK$ and $gJ$, all have the absolute dimension of an
action $h$.  In particular,
\begin{equation}
  [g]=\frac{h}{q}\,,\quad [e]=q\quad\Longrightarrow\quad [g]\times
  [e]=h\,.
\end{equation}
The ultimate quantized nature of the action then clearly indicates
that the product of the electric and magnetic charges should be an
integer multiple of $h$. Such a qualitative conclusion, based on sheer
dimensional analysis within the premetric approach, appears to be in
complete agreement with the old observation of Bialynicki-Birula
\cite{BB71a,BB71b} which shows, in essence, that the 2-form $eF - gH
=$ {\it exact 2-form} $+$ a possible $\delta$-like 2-form. Taking the
exterior derivative of this relation and using the generalized Maxwell
equations (\ref{newmaxwell}), one then arrives at the quantization
condition of the electric and magnetic charge in the form of Schwinger
and Zwanziger, see \cite{BB71a}.

The complex electromagnetic field $U$ represents a 4-dimensional
version of the Riemann-Silberstein fields. It provides a compact
formulation of the electrodynamic field equations with and without
magnetic charges and demonstrates the primary character of the
metric-free reciprocity operator as compared to the metric-dependent
Hodge duality operator.

{}From a mathematical point of view, the modified axiomatics proposed
here is consistent and it naturally generalizes the approach developed
earlier in \cite{HO3}. However, we should honestly mention that some
potentially serious physical problems may arise in this framework. In
the standard formulation, electromagnetism manifests itself in two
fundamentally different physical quantities: excitation and field
strength. In simple terms, they represent the answers to the questions
``How many?" and ``How strong?", respectively. When we allow for
magnetic charge, this important and clear-cut physical distinction of
quantities and intensities is lost, which may lead to a possible
confusion in the interpretation of the electromagnetic fields $H$ and
$F$. When combined with the lack of any experimental evidence for
monopoles \cite{Abbott,He,Kalbfleisch}, such a theoretical problem
makes us skeptical about the existence of magnetic charges in nature.

\subsection*{Acknowledgments} 
We thank Gerry Kaiser and Bahram Mashhoon for stimulating discussions.
Our work has been supported by the DFG project HE 528/20-1.

\end{document}